\documentclass[times, twoside]{zHenriquesLab-StyleBioRxiv}

\leadauthor{Song} 

\begin{document}

\title{Challenges and Lessons from MIDOG 2025: A Two-Stage Approach to Domain-Robust Mitotic Figure Detection}
\shorttitle{Approach for MIDOG 2025}

\author[1]{Euiseop Song}
\author[2]{Jaeyoung Park}
\author[3]{Jaewoo Park}

\affil[1]{Department of Medicine, Korea University Graduate School, South Korea}
\affil[2]{Department of Medicine, Dongguk University, South Korea }
\affil[3]{Department of Computer Science and Engineering, Sogang University, South Korea }

\maketitle

\begin{abstract}
Mitotic figure detection remains a challenging task in computational pathology due to domain variability and morphological complexity. This paper describes our participation in the MIDOG 2025 challenge, focusing on robust mitotic figure detection across diverse tissue domains. We developed a two-stage pipeline combining Faster R-CNN for candidate detection with an ensemble of three classifiers (DenseNet-121, EfficientNet-v2, InceptionResNet-v2) for false positive reduction. Our best submission achieved F1-score 0.2237 (Recall: 0.9528, Precision: 0.1267) using a Faster R-CNN trained solely on MIDOG++ dataset. While our high recall demonstrates effective mitotic figure detection, the critically low precision (12.67\%) reveals fundamental challenges in distinguishing true mitoses from morphologically similar imposters across diverse domains. Analysis of six submission variants showed that subsequent optimization attempts were counterproductive, highlighting the complexity of domain generalization in histopathology. This work provides valuable insights into the practical challenges of developing robust mitotic figure detection algorithms and emphasizes the importance of effective false positive suppression strategies. 
\end{abstract}


\begin{corrauthor}
proproscrinator31@gmail.com
\end{corrauthor}

\section*{Introduction}

Mitotic figure detection represents one of the most critical yet challenging tasks in computational pathology. The mitotic count serves as a fundamental prognostic biomarker for tumor grading across various cancer types, including breast carcinoma, where it constitutes an essential component of the Nottingham histologic grade. However, accurate mitotic figure identification presents significant challenges even for experienced pathologists due to several factors: the small size of mitotic figures relative to the tissue context, morphological variability across different mitotic phases, and the presence of mitotic-like figures (imposters) that closely resemble true mitoses.

The inherent subjectivity in mitotic figure recognition has been quantitatively demonstrated in large-scale studies. The 1000 Mitoses Project\cite{Lin2024-jr}, a consensus-based international collaborative study, evaluated 1,010 mitotic figures from TCGA datasets with 85 pathologists, revealing a median agreement rate of 80.2\% with substantial variability (range: 42.0\%-95.7\%). This inter-observer variability significantly impacts the reproducibility and clinical utility of mitotic counts, highlighting the urgent need for robust automated detection systems.

The MItosis DOmain Generalization (MIDOG) challenge series, initiated in 2021, addresses these limitations by focusing on scanner-agnostic and domain-robust mitotic figure detection algorithms. The challenge has evolved from single-domain breast cancer detection (MIDOG 2021)\cite{midog2021} to multi-domain generalization across tumor types and species (MIDOG 2022)\cite{midog2022}, establishing comprehensive benchmarks for algorithm evaluation. MIDOG 2025 extends this scope with two distinct tracks: Track 1 focuses on robust mitotic figure detection across adverse tissue conditions, while Track 2 introduces the clinically important task of atypical mitotic figure classification.

Current state-of-the-art approaches, exemplified by the MIDOG 2022 winning algorithm "Mitosis Detection, Fast and Slow" (MDFS), employ two-stage pipelines that first identify candidate regions using segmentation models and subsequently classify these candidates using patch-based classifiers. This approach effectively balances sensitivity in candidate detection with precision in final classification, addressing the inherent class imbalance in histopathological images.

Building upon these established principles, we developed a two-stage detection pipeline combining object detection for candidate identification with ensemble classification for false positive reduction. Our approach aims to achieve robust performance across the diverse domains represented in MIDOG 2025 while maintaining computational efficiency suitable for clinical deployment.

\section*{Material and Methods}

\begin{figure}
\centering
\includegraphics[width=\linewidth]{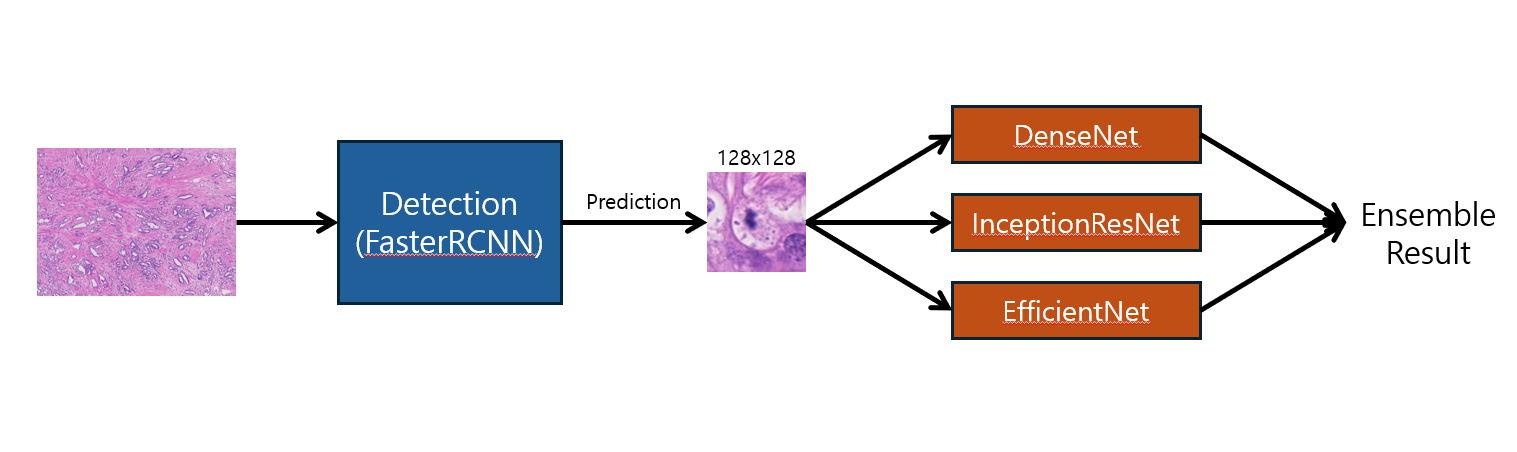}
\caption{Overview of our full pipeline}
\label{fig:computerNo}
\end{figure}

\textbf{A. Detection }For detector training, we utilized a pooled dataset (PooledDB) combining MIDOG++, MITOS-CMC, and MITOS-CCMCT datasets. The PooledDB comprises 503 large regions of interest (ROIs) and 54 whole slide images (WSIs) spanning 8 tumor types from human and canine specimens, providing diverse morphological and staining variations essential for robust mitotic figure detection.

We implemented a Faster R-CNN architecture as the detection backbone. Training patches of 512×512 pixels were randomly generated based on annotations with a sampling ratio of 5:1:4 for foreground, random, and imposter patches, respectively. This sampling strategy was designed to balance positive examples with challenging negative cases that resemble mitotic figures. The training set comprised 4,096 patches with 1,024 validation patches, using an NMS threshold of 0.4 and training for 150 epochs.

Our preprocessing pipeline included D4 transformations for rotation invariance, Defocus augmentation (radius 1-3, p=0.3), and RandStainNA\cite{Shen_2022} for stain normalization. RandStainNA performs quantitative analysis of stain characteristics in color space to generate appropriate stain style templates, enabling realistic stain augmentation. We generated color profiles based on the MIDOG++ dataset with parameters std\_hyper=-0.7 and p=0.3 to account for inter-laboratory staining variations.

The detection model evaluation on PooledDB test set revealed significant challenges: while achieving high recall (0.9820), precision was critically low (0.0578), resulting in F1-score of 0.1091. This indicates substantial false positive generation, suggesting the need for improved discrimination between true mitotic figures and morphologically similar structures.

\textbf{B. Classification} We designed an ensemble classification approach using three complementary backbone architectures: DenseNet-121\cite{huang2018denselyconnectedconvolutionalnetworks}, EfficientNet-v2\cite{tan2021efficientnetv2smallermodelsfaster}, and InceptionResNet-v2\cite{szegedy2016inceptionv4inceptionresnetimpactresidual}. Due to time constraints, models were trained on different datasets with varying protocols, which may have introduced methodological inconsistencies.

\textit{EfficientNet-v2} was trained using PooledDB-derived patches. We extracted 128×128 pixel patches centered on annotations and detection-generated bounding boxes, resulting in 583,902 training patches (123,614 mitotic, 460,288 non-mitotic) and 99,463 validation patches (20,790 mitotic, 78,673 non-mitotic). Training parameters included: batch size 32, 20 epochs, learning rate 1e-4, Adam optimizer, ReduceLROnPlateau scheduler, and CrossEntropyLoss. The model achieved test accuracy of 0.9327 with mitotic figure F1-score of 0.83 (precision: 0.85, recall: 0.82).

\textit{InceptionResNet-v2} was trained on combined MIDOG++, MiDeSeC, and ICPR 2012 datasets (14,681 mitotic, 12,068 non-mitotic patches). Initial training with FocalLoss and WeightedRandomSampler resulted in gradient explosion, likely due to unnecessary imbalance correction given the relatively balanced dataset. We subsequently employed AdamW (lr=1e-4), batch size 32, 50 epochs, cosine scheduling with 5-epoch warmup, mixed precision, early stopping based on validation F1, CrossEntropyLoss, and RandomSampler, achieving accuracy 0.810 and F1-score 0.806.

\textit{DenseNet-121} initially used identical settings to InceptionResNet-v2 but showed limited improvement. We applied knowledge distillation\cite{hinton2015distillingknowledgeneuralnetwork} using InceptionResNet-v2 as teacher model (temperature=4.0, loss ratio 0.5 for student CrossEntropy and KD components). Additional techniques included EMA (decay=0.9998) and color augmentation (brightness=0.1, contrast=0.1, saturation=0.05, hue=0.03, random grayscale p=0.05, Gaussian blur p=0.05). Knowledge distillation significantly improved performance from F1-score 0.57 to 0.73.

\textbf{C. Full Pipeline }Our pipeline processes detection-generated bounding boxes by extracting 128×128 patches for ensemble classification, followed by mitotic figure filtering. The ensemble averages softmax probabilities from all three models.

Evaluation revealed a critical performance bottleneck: while detection alone achieved F1-score 0.0831 (recall 0.9842), the full pipeline degraded to F1-score 0.0646 (recall 0.0488). This suggests that the classification stage, while reducing false positives, inadvertently filtered out numerous true positives, highlighting the need for better integration between detection and classification components and more sophisticated ensemble calibration strategies.

\section*{Results}

Six algorithmic variants were submitted to MIDOG 2025, evaluating different detection model and classifier combinations. Our best-performing submission employed a Faster R-CNN trained solely on MIDOG++, achieving F1-score 0.2237, Recall 0.9528, and Precision 0.1267.

Our final optimization attempt submitted before September 1st yielded F1-score 0.2031, Recall 0.8055, and Precision 0.1162. This 9.2\% decline in F1-score from our initial submission indicates that subsequent dataset restructuring and classifier tuning efforts were counterproductive.

The consistently low precision across all submissions (0.1162-0.1267) reveals a critical limitation in false positive suppression, while high recall values (0.8055-0.9528) confirm effective mitotic figure detection. The performance deterioration suggests that our optimization attempts introduced overfitting, reducing true positive detection without meaningful improvements in specificity. These results highlight the challenge of distinguishing mitotic figures from morphologically similar imposters across diverse tissue domains.

\section*{Discussion}

This work represents our first algorithmic challenge in digital pathology, focusing on mitotic figure detection across diverse domains. While our final performance was below competitive levels, we gained valuable experience in experimental design, deep learning backbone selection for pathology applications, and the critical importance of systematic logging and pipeline development for reproducible research.

We plan to systematically investigate the overfitting mechanisms through targeted ablation studies focusing on regularization techniques and domain adaptation strategies. Our immediate goal is to develop improved models validated on controlled in-house datasets with reduced stain variation, establishing a foundation for more robust algorithms capable of handling diverse pathological imaging environments.

\begin{acknowledgements}
This paper was written in Korean and translated into English by support of \textit{Claude Sonnet 4}.
\end{acknowledgements}

\section*{Bibliography}
\bibliography{literature}

\end{document}